\documentclass[runningheads,a4paper]{llncs}
\usepackage{amssymb} 
\usepackage{graphicx}
\usepackage{url}
\usepackage{float}
\usepackage{caption}

\usepackage[english]{babel}

\usepackage[mag=1000,a4paper,left=3.5cm,right=3.5cm,top=3cm,bottom=3cm]{geometry} 
\usepackage{amsmath}
\usepackage{listings}   
\usepackage{verbatim}
\usepackage{todonotes}
\usepackage{xcolor}
\usepackage{url}

\lstdefinelanguage{Jolie}{
morekeywords={provide,until,OneWay,RequestResponse,new, type, main,define,inputPort,outputPort,init,execution,include,	cset,if,else,csets,interface, throws,global,constants,for,
foreach,while,int,double,raw,void,undefined,string,long,bool,any,single, sequential, concurrent, Jolie, Java, JavaScript, embedded, Location, Protocol, Interfaces, Aggregates, scope, install, cH, comp, throw, this, default, synchronized, nullProcess, false, true},
sensitive=true,
comment=[l]{//},
morecomment=[s]{/*}{*/},
morestring=[b]",
otherkeywords={;,|,:}
}

\lstdefinelanguage{z3}{
morekeywords={assert, false, true, declare, sort, fun, Bool, String, RegEx, forall, const, define, let, not, implies, iff, and, or},
sensitive=true,
comment=[l]{//},
morecomment=[s]{/*}{*/},
comment=[l]{;},
morestring=[b]",
otherkeywords={|,:}
}

\begin{document}

\title{Jolie Static Type Checker: a prototype}

\authorrunning{Daniel de Carvalho et al.}

\author{
Daniel de Carvalho\inst{1},
Manuel Mazzara \inst{1},
Bogdan Mingela\inst{1} \\
Larisa Safina \inst{1},
Alexander Tchitchigin \inst{2},
Nikolay Troshkov \inst{1}
\institute{Innopolis University, Russia 
\and Typeable.io LLC, Russia}}

\maketitle

\begin{abstract}
Static verification of a program source code correctness is an important element of software reliability. Formal verification of software programs involves proving that a program satisfies a formal specification of its behavior. Many languages use both static and dynamic type checking. With such approach, the static type checker verifies everything possible at compile time, and dynamic checks the remaining. The current state of the Jolie programming language includes a dynamic type system. Consequently, it allows avoidable run-time errors. A static type system for the language has been formally defined on paper but lacks an implementation yet. In this paper, we describe a prototype of Jolie Static Type Checker (JSTC), which employs a technique based on a SMT solver. We describe the theory behind and the implementation, and the process of static analysis.
\end{abstract}

\section{Introduction}

The microservice architecture is a style inspired by service-oriented computing that promises to change the way in which software is perceived, conceived and designed \cite{mustafin2017}. The trend of migrating monolithic architectures into microservices to reap benefits of scalability is growing fast today \cite{Dragoni2017b,DDLM2017}. Jolie \cite{MGZ14} is the only language natively supporting microservice architectures \cite{Guidi2017} and, currently, has dynamic type checking only.

Static type checking is generally desirable for programming languages improving software quality, lowering the number of bugs and preventing avoidable errors. The idea is to allow compilers to identify as many issues as possible before actually run the program, and therefore avoid a vast number of trivial bugs, catching them at a very early stage. Despite the fact that, in the general case interesting properties of programs are undecidable \cite{Rice53}, static type checking, within its limits, is an effective and well established technique of program verification. If a compiler can prove that a program is well-typed, then it does not need to perform dynamic safety checks, allowing the resulting compiled binary to run faster. 

A static type system for Jolie has been exhaustively and formally defined only on paper \cite{nielsen}, but still lacks an implementation. The obstacles of programming in a language without a static type analyzer have been witnessed by Jolie developers, especially by newcomers. However, implementing such system is a non trivial task due to technical challenges both of general nature and specific to the language. In this paper, we introduce and describe the Jolie Static Type Checker (JSTC), building on top of the previous work on the Jolie programming language \cite{MGZ14}. Our approach follows the formal derivation rules as defined in \cite{nielsen}. The project is built as a Java implementation of source code processing and verification via Z3 SMT solver \cite{MouraB08} and it has to be intended as our community contribution to the Jolie programming language \cite{Bandura16}.  

Section~\ref{background} recalls the basic of Jolie and section~\ref{related-work} discusses related work. The description of the static type-checking and the system architecture can be found in Section~\ref{impl}, while Section~\ref{conclusions} draws conclusive remarks and discusses open issues.

\section{Background}
\label{background}


Microservices~\cite{Dragoni2017} is an architectural style evolved from Service-Oriented Architectures~\cite{mackenzie2006}. According to this approach, applications are composed by small independent building blocks that communicate via message passing. These composing parts are indeed called microservices. This paradigm has  seen a dramatic growth in popularity in recent years~\cite{N15}.
Microservices are not limited to a specific technology. Systems can be built using a wide range of technologies and still fit the approach. In this paper, however, we support the idea that a paradigm-based language would bring benefit to development in terms of simplicity and development cost. 

Jolie is the first programming language constructed above the paradigm of microservices: each component is autonomous service that can be deployed separately and operated by running in parallel processes. Jolie comprises formally-specified semantics, inspired by process calculi such as CCS~\cite{milnerCCS} and the $\pi$-calculus~\cite{Milner1999}. As for practical side, Jolie was inspired by standards for Service-Oriented Computing such as WS-BPEL~\cite{BPEL} and the attempts of formalizing it \cite{Mazzara:phd}. The composition of both theoretical and practical aspects allows Jolie to be the preferred candidate for the application of modern research methodologies, e.g. runtime adaptation, process-aware web applications, or correctness-by-construction of concurrent software.

The basic abstraction unit of Jolie is the microservice~\cite{Dragoni2017}. It is based on a recursive model where every microservice can be easily reused and composed for obtaining, in turn, other microservices. Such approach allows distributed architecture and guarantees simple management of all components, which reduces maintenance and development effort. Microservices communicate and work together by sending messages to each other. In Jolie, messages are represented in tree structure. A variable in Jolie is a path in a data tree and the type of a data tree is a tree itself. Equality of types must therefore be handled with that in mind. Variables are not declared, wherefore the manipulation of the program state must be inferred. Communications are type checked at runtime, when messages are sent or received. Type checking of incoming messages is especially relevant, since it could moderate the consequences of errors. 

The Jolie language is constructed in three layers: The behavioural layer operates with the internal actions of a process and the communication it performs seen from the process point of view, the service layer deals with the underlying architectural instructions and the network layer deals with connecting communicating services.

Other workflow languages are capable of expressing orchestration of (micro)services the same way Jolie can do, for example WS-BPEL~\cite{BPEL}.  WS-BPEL allows developers to describe workflows of services and other communication aspects (such as ports and interfaces), and it has been also shown how dynamic workflow reconfiguration can be expressed \cite{MazzaraADB11}. However, WS-BPEL has been designed for high-level orchestration, while  programming the internal logic of a single micro-service requires fine-grained procedural constructs. Here it is were Jolie works better.

\section{Related work} \label{related-work}

The implementation of a static type checker for Jolie is part of a broader attempt to enhance the language for practical use. Previous work on the type system has been done, however focusing mostly on dynamic type checking. Safina extended the dynamic type system as described in~\cite{Safina2016}, where type choices have been added in order to move computation from a process-driven to a data-driven approach.

The idea to integrate dynamic and static type checking with the introduction of refinement types, verified via SMT solver, has been explored in~\cite{Tchitchigin16}. The integration of the two approaches allows a scenario where the static verification of internal services and the dynamic verification of (potentially malicious) external services cooperates in order to reduce testing effort and enhancing security.

The idea of using SMT Solvers for static analysis, in particular in combination with other techniques, has been successfully adopted before for other programming languages, for example LiquidHaskell and F*. LiquidHaskell \cite{LiquidHaskell}\footnote{Online demo at \url{http://goto.
ucsd.edu/~rjhala/liquid/haskell/demo/}} is a notable example of implementation of 
Liquid Types (\textit{Logically Qualified Data Types}) \cite{Rondon2008}. It is a 
static verification technique combining \textit{automated deduction} (SMT solvers), \textit{model checking} (Predicate Abstraction), and \textit{type systems} (Hindley-Milner inference). Liquid Types have been implemented for several other programming languages. The original paper  presented an OCaml implementation. F* \cite{F*} instead an ML-like functional programming language specifically designed for program verification.The F* type-checker uses a combination of SMT solving and manual proofs to guarantee correctness

Another direction in developing static type checking for Jolie is creating the verified type checker\footnote{\url{https://github.com/ak3n/jolie}} by means of proof assistant instead of SMT solver \cite{akentev2017}. Proof assistant is a software tool needed to assist with the development of formal proofs by human-machine collaboration and helps to ascertain the correctness of them. The type checker is expressed as well-typed program with dependent types in Agda~\cite{agda}. If the types are well formed, all required invariants and properties are described and expressed in the types of the program meaning that the program is correct. This work is currently in progress and evolves in parallel with ours. 

\section{Static type-checking implementation}\label{impl}

This paper builds on top of Julie Meinicke Nielsen's work~\cite{nielsen} at the Technical University of Denmark implementing the type system of the Jolie language. The thesis represents the theoretical foundation for the type checking of the core fragment of the language, which excludes recursive types, arrays, subtyping of basic types, faults and deployment instructions such as architectural primitives. The work of Nielsen presents the first attempt at formalizing a static type checker for the core fragment of Jolie, and the typing rules expressed there are the core theory behind our static checker.

In Nielsen's work typing rules are represented in the style of type theory where type rules are inference rules describing how a type system assigns a type to a syntactic construct of the language \cite{Cardelli:1996}. The rules are then applied by the type system to determine if a program is well typed or not. The main typing rules will be presented in the following of this paper.

The implementation of JSTC consists of two system components. Firstly, a Java program accepts the source code of a Jolie program, builds an abstract syntax tree (AST), visits it and produces a set of logical assertions written in SMT Lib~\cite{smtspec} language. At the second phase, the generated assertions are feed into Z3 solver. The basic idea is to implement, for each Jolie node\footnote{Any syntax unit is considered a node. It can be a logical or arithmetic expression, an assignment; a condition; a loop etc. Those nodes comprise the abstract syntax tree.}, methods containing statements expressed in the SMT Lib syntax. These statements can then be processed via a solver. In Figure 1 the overall process is pictorially represented and details are described in section~\ref{impl:translation}.

\begin{figure}
\label{dataflow}
\centering
\includegraphics[width=80mm]{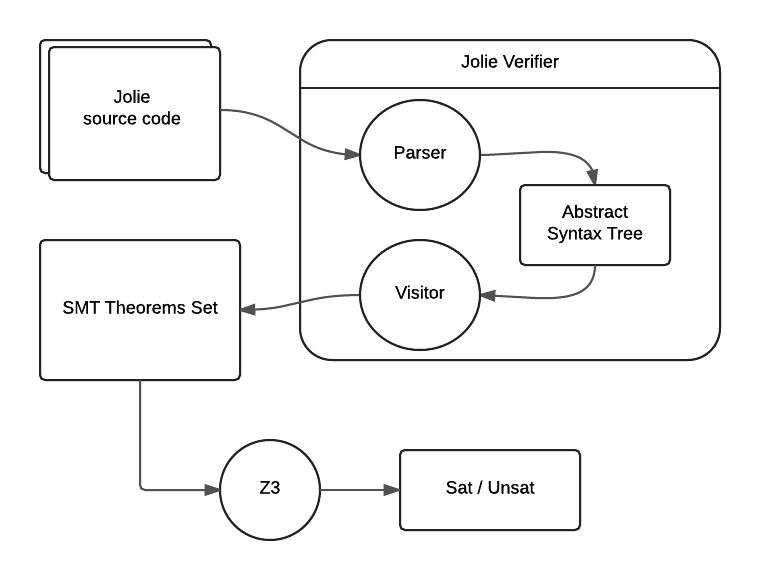}
\caption{Process of Type Checking in the Jolie Verifier}
\end{figure}

The concept of SMT solvers is closely related to logical theorems. Logic, especially in the field of proof theory, considers theorems as statements of a formal language. Existence of such logical expressions allows to formulate a set of axioms and inference rules to formalize the typing rules for each of Jolie syntax nodes and then perform the validation of the nodes using constructed theorems. Consequently, the Jolie typing rules are the specific cases of logical theorems, that are used in the project.  The concept is implied from software verification fundamentals~\cite{Hoare}. 

Since Jolie program may contain complex expressions with function calls, it is also necessary to consider data structures representing a match between names and expressions, in order to be able to avoid inconsistency and redundancy, that are likely to cause conflicts during type-checking. The project implementation considers using a stack during the recursive checking of the nodes as illustrated in section~\ref{impl:translation}.

The decision of using an SMT-solver, instead of more lightweight techniques, was made in order to allow a future straightforward integration of Refinement Types into the type checker, objective on which our team is already working \cite{Safina2016,Tchitchigin16}. Furthermore, relying on a solid existing technology allowed us to prototype and release a proof of concept of the type checker in a shorter period of time.

\subsection{Jolie verifier}\label{impl:parser}

The Java program reuses an existing structure of a Visitor pattern that was used in a previous project for formatting Jolie source code \footnote{\url{https://github.com/nickaleks/jolie}}. It accepts processed Jolie program source code in the form of AST and performs traversing. For each kind of node the system creates one or more logical formulas written using SMT-LIB \cite{smtspec} syntax, which are then stored into a file on disk. {At the current implementation state the theorems are collected in a single data element.}The verifier targets assignments, conditions, and other cases of variables usage where type consistency can be violated.

\subsection{SMT Solver}\label{impl:smt}

Z3 carries out the main functionality of program verification. Z3 is an SMT solver from Microsoft Research~\cite{MouraB08}. It is targeted at solving problems that arise in software verification and software analysis. Given a set of formulas that was previously created by the verifier in Java, Z3 processes it and returns whether this set is satisfiable or not. In case of any contradiction in the set, the solver will signal that the overall theorem is not satisfiable, therefore alerting that the input program is not consistent in terms of types usage.

\subsection{Typing rules}\label{impl:trules}

Our objective is to accurately translate Jolie typing rules into SMT statements, therefore allowing static type checking \footnote{Please note that, at the moment, not all the rules in ~\cite{nielsen} have been implemented.}. The foreground activity so far is producing the set of statements for the construct of the behavioural layer of Jolie. The layer describes the internal actions of a process and the communications it performs seen from the process' point of view. The layer is chosen for the first phase of the development because of being the foundation of the syntactical structures of Jolie. Also there is a similarity of the layer with common programming languages in a sense of the abstraction level. So these facts make the behavioural level to be the first entry in the world of Jolie language capabilities.

All statements at the behavioural layer of Jolie are called behaviours. We write $\Gamma \ \vdash_{B} \space B \triangleright  \Gamma'$ to indicate a behaviour $B$, typed with respect to an environment $\Gamma$, which updates $\Gamma$ to $\Gamma'$ during type checking~\cite{nielsen}. 

There are some core rules presented and described below:

\textbf{T-Nil.} The typing rule for a nil behaviour is an axiom. In the conclusion the typing environment is not changed, since the \textrm{nil} statement doesn't affect the typing environment.

\begin{center}
$\dfrac{}{\Gamma \ \vdash_{B} 0 \triangleright  \Gamma}$
\end{center}
\normalsize

\textbf{T-If-Then-Else.} The rule for typing an if statement is standard: An if statement is typable if its condition has type \textrm{bool}, and if the type checking of its branches perform the same updates to the environment. We require the branches to perform the same updates because we do not know which branch will be taken. The else part may also be omitted and \textrm{B2} may be replaced by an empty behaviour. The conditional typing statement is the following:

\begin{center}
$\dfrac{\Gamma \ \vdash \ e:bool \quad \Gamma \ \vdash_{B} B_1 \triangleright  \Gamma' \quad \Gamma \ \vdash_{B} B_2 \triangleright  \Gamma'}{\Gamma \ \vdash_{B} if(e) \ B_1 \ else \ B_2 \triangleright  \Gamma'}$
\end{center}
\normalsize

\textbf{T-While.} The rule for typing a \textrm{while} statement is standard: A while statement is typable if its condition has type \textrm{bool}, and if type checking its body has no influence on the typing environment.

\begin{center}
$\dfrac{\Gamma \ \vdash \ e:bool \quad \Gamma \ \vdash_{B} B \triangleright  \Gamma}{\Gamma \ \vdash_{B} while(e) B \triangleright  \Gamma}$
\end{center}
\normalsize


Above, it is required that the body of the while loop does not change the typing of variables because we do not know whether the body will be executed at all, and for how many times. We also require that expression \textrm{e} is type checked against type \textrm{bool}.

\textbf{T-Seq.} A sequence statement typed with respect to an environment is typable if its first component is typable with respect to the environment and its second component is typable with respect to the update of the environment performed by the first component. The update of the environment performed by the sequence statement is the update performed by the second component with respect to the update performed by the first component.

\begin{center}
$\dfrac{\Gamma \ \vdash_{B} \ B_1 \triangleright  \Gamma' \quad \Gamma' \ \vdash_{B} B_2 \triangleright  \Gamma''}{\Gamma \ \vdash_{B} B_1;B_2 \triangleright  \Gamma''}$
\end{center}
\normalsize

Thus, fundamental typing rules of the behavioral layer of Jolie programming language are presented and explained for further topic revelation.

\subsection{Typing rules to SMT translation}\label{impl:translation}
Here we will illustrate an example of the conditional rule translation in order to understand the procedure in detail.





The typing rule of the \textit{if} statement does not contradict intuition. The statement is typeable when its condition expression is boolean, and the execution of both its branches brings the same updates to the environment. This means that the set of matches between expressions and variables with their types remains the same with no difference from a branch choice. This is necessary since it is not possible to predict what branch will be executed at runtime \footnote{The \textit{else} part may also be omitted and $B_2$ may be replaced by an empty behavior.}. 

The full implementation is available on github.\footnote{\url{https://github.com/innopolis-jolie-smt-typechecker/jolie}} Below we show the Java fragment that builds the corresponding SMT statement.



\begin{verbatim}
public void visit(IfStatement n) {
   for (Pair<OLSyntaxNode, OLSyntaxNode> statement : n.children()) {
      OLSyntaxNode condition = statement.key(); 
      OLSyntaxNode body = statement.value();
      check(condition); 
      TermReference conditionTerm = usedTerms.pop(); 
      writer.writeLine("(assert (hasType " + conditionTerm.id + " bool))");
      if (body != null) {body.accept(this);}
   } 
   if (n.elseProcess() != null){n.elseProcess().accept(this);}
}
\end{verbatim}

The code structure represents basic steps to achieve a record with corresponding SMT statements of the block as a result. Firstly, a condition of the \textit{if} statement is separated from the body. Then the condition is sent to be checked using the same visitor class. Eventually after the last 'recursion' step the condition is put in the stack of terms, which contains any terms (expressions, variables etc.) processed during the checking. So the term corresponding to the condition is expected to be on top of the stack. Then an assertion that says the condition term is boolean is written. Afterwards the body is processed using one of the other overloads of the visitor. These steps can be repeated in case of existence of nested conditional statements. In the end of the method the \textit{else} branch body of the very first \textit{if} is processed if it is present. There is also an important note is that the conditional statement does not impose any other direct type restrictions besides the condition term that is confirmed by the mentioned typing rule. Other implemented nodes can be seen in the source mentioned above.

The Jolie verifier takes some input for processing. Let us consider a simple piece of Jolie code with a conditional statement.


\begin{verbatim}
a = 2;
b = 3;
if ( a > b ) {
   println@Console( a + b )()
} else{
   println@Console( "Hello!" )()
}
\end{verbatim}

In the case everything works, none of the typing rules is violated. Z3 agrees with the opinion and results in 'sat', that means the program state is satisfiable. We list here the SMT statements representing the condition processing:

\begin{verbatim}
(declare-const $$__term_id_10 Term)
(assert (hasType $$__term_id_10 bool))

(assert (hasType $$__term_id_10 bool))
\end{verbatim}





The first assertion is made based on an expression type determination: the expression $a > b$ is boolean. The second one is imposed by the typing rule: the condition expression must be boolean. In this case there is no contradiction between these two assertions.

If the condition would be replaced with some other type expression the typing rule may be violated. The corresponding example case with a replacement of $a > b$ is shown below:

\begin{verbatim}
a = 2;
b = 3;
if ( 5 ) {
   println@Console( a + b )()
} else{
   println@Console( "Hello!" )()
}
\end{verbatim}

And the constructed SMT statements for the condition expression are given here:

\begin{verbatim}
(declare-const $$__term_id_10 Term)
(assert (hasType $$__term_id_10 int))

(assert (hasType $$__term_id_10 bool))
\end{verbatim}

Now the contradiction between the assertions is notable. The parser decided the expression to be an integer, which is correct. But the restriction on a condition type from the typing rule simply contradict with the actual type. Consequently Z3 results in 'unsat'. This means that the program state representing the assertion unsatisfiable and incorrect in terms of the considered static type checking analysis.

\section{Evaluation} \label{evaluation}

The question of how to prove correctness of verification tools has always been widely discussed. How can we be sure that the output of such tool is correct? It can be poorly written, or the hardware could malfunction. However, in most cases we tend to trust verification tools, and in our project we have to make sure that this tool is as trustworthy as any other. The general solution is testing. Verification of the written code correctness was continuously performed during the development process. The Jolie Team created a collection of examples of Jolie programs.\footnote{\url{https://github.com/jolie/examples}} The verification results of some of these are presented in this section.

\subsection{An unsatisfiable model} \label{evaluation:cases:unsat}

The general purpose of the type checker is to find inconsistency in types usage. The program listed below is the most basic example of a program with inconsistent types. The variable \textit{myInt} is assigned an integer first, and then a string. The current design of the type checker disallows this behavior.


\begin{verbatim}
main 
{
   myInt = 15;
   myInt = "fifteen"
}
\end{verbatim}


The resulting set of SMT theorems is listed below.

\begin{verbatim}
(declare-const myInt Term)

(declare-const $$__term_id_4 Term)
(assert (hasType $$__term_id_4 int))
(assert (sameType myInt $$__term_id_4))
(assert (hasType myInt int))

(declare-const $$__term_id_10 Term)
(assert (hasType $$__term_id_10 string))
(assert (sameType myInt $$__term_id_10))
(assert (hasType myInt string))
\end{verbatim}

The model is unsatisfiable. Assertions on the lines 4-6 restrict type of \textit{myInt} to integer, whereas assertions on the lines 9-11 ensure that the same variable should be of type string. These assertions cannot be evaluated to be true in the same model, considering the initial theorems of the type checker model. Therefore, the overall model is going to be unsatisfiable upon calling the \textit{check-sat} Z3 command.

\subsection{A satisfiable model} \label{evaluation:cases:sat}

In case everything in the program code is correct in terms of type consistency, the type checker should evaluate the resulting SMT model as satisfiable. It also should ignore cases that have not being processed properly yet, without giving any false positives on any inconsistency. The program listed below has, in fact, an inconsistency in types usage. The line 8 reassigns a variable to be of type integer, whereas at the line 5 the same variable was introduced as a variable of type string. However, as long as this assignment include a statement with a dynamic key, the type checker ignores it. The reason for this is inability to determine which variable this variable path will point to at the moment of execution.


\begin{verbatim}
main 
{
  key = "cat";
  animals.cat = "I am a cat";
  animals.(key) = 13
}
\end{verbatim}

The resulting set of SMT theorems is listed below. 

\begin{verbatim}
(declare-const key Term)

(declare-const $$__term_id_4 Term)
(assert (hasType $$__term_id_4 string))
(assert (sameType key $$__term_id_4))
(assert (hasType key string))
(declare-const $$__term_id_5 Term)
(declare-const animals Term)
(declare-const animals.cat Term)

(declare-const $$__term_id_10 Term)
(assert (hasType $$__term_id_10 string))
(assert (sameType animals.cat $$__term_id_10))
(assert (hasType animals.cat string))
(declare-const $$__term_id_11 Term)
(declare-const animals.DYNAMIC_PATH_$$__term_id_14 Term)

(declare-const $$__term_id_19 Term)
(assert (hasType $$__term_id_19 int))
(assert (sameType animals.DYNAMIC_PATH_$$__term_id_14 $$__term_id_19))
(assert (hasType animals.DYNAMIC_PATH_$$__term_id_14 int))
\end{verbatim}

The workaround for the dynamic keys issue can be seen at Lines 20 and 21. Any time a variable is used in order to construct another variable path, the whole Term is getting a unique identifier marked with the \textit{DYNAMIC\_PATH} substring. This way it will not interfere with any other theorem, thus not affecting the overall model satisfiability.


\section{Conclusions and future works} \label{conclusions}

Jolie is the first programming language specifically oriented to the microservice architecture. It has been shown how software attributes such as extensibility, modifiability and consistency can significantly benefit from a migration into the microservice paradigm~\cite{Dragoni2017b,DDLM2017}. Projects run by our team demonstrated the efficacy of the paradigm and of the Jolie programming language in the field of ambient intelligence and smart buildings~\cite{Salikhov2016a,Salikhov2016b}. Social networks implementation would also benefit from a reorganization of the software architecture~\cite{MBGDMQN2013}. Local projects, and beyond that a number of projects worldwide involving the use of Jolie, would immensely benefit from a fully stable implementation of the Jolie Static Type Checker.

Static type checking allows compilers to identify certain programming mistakes (that violate types) at compile time, i.e. before actually running the program. Therefore a vast number of trivial bugs can be caught and fixed at a very early stage of the software life-cycle. In this paper we described JSTC, a static type checker for the Jolie programming language which natively supports microservices. A static type system for the language has been exhaustively and formally defined on paper, but so far still lacked an implementation. We introduced our ongoing work on a static type checker and presented some details of the implementation. The type checker prototype, at the moment, consists of a set of rules for the type system expressed in SMT Lib language. The actual implementation covers operations such as assignments, logical statements, conditions, literals and comparisons.

JSTC is already able to validate programs, as it has been shown in this paper. However, it works with certain assumptions. The main assumption is that programs do not contain implicit type casts. The Jolie language allows implicit type casts, however, their behavior is very complex. Handling such situations is an open issue left for future development and future versions. Two other major issues have not been addressed.

\paragraph{Variable types can be changed at runtime.} This strictly depends on the approach that has been chosen. Generally, static typing guarantees that a variable has a type that cannot be changed after declaration or assignment. However, Jolie allows this operation. We need to determine which behavior we expect from the type checker, thus deciding how to process type changes.

\paragraph{Implicit type casts in Jolie are ambiguous.} This is a major problem, and  further research is required in order to find a solution. While Jolie allows implicit type casts, sometimes the result of a cast is not obvious. For example, casting a negative Integer to Boolean will result in a False. This is an unexpected behavior when compared to other programming languages. There may be a solid rationale for this, however, we need to investigate all cases and make sure that the type checker works accordingly to the Jolie actual behavior, and not to the expected one.

JSTC future releases will need to be validated in real-life applications. The plan is to use the Jolie programming language and the type checker as a basis for the development of future research projects, the same way was done in \cite{Salikhov2016a} and \cite{Salikhov2016b}. Potential application scenarios are cognitive architecture \cite{Vallverdu16}, automotive systems \cite{Gmehlich2013} and smart houses \cite{Nalin2016}.

\bibliographystyle{plain}
\bibliography{references}

\begin{thebibliography}{10}

\bibitem{F*}
{F*}.
\newblock \url{https://www.fstar-lang.org/ }.

\bibitem{BPEL}
{WS-BPEL} {OASIS} {W}eb {S}ervices {B}usiness {P}rocess {E}xecution {L}anguage.
  accessed {April} 2016.
\newblock
  \url{http://docs.oasis-open.org/wsbpel/2.0/wsbpel-specification-draft.html}.

\bibitem{akentev2017}
Evgenii Akentev, Alexander Tchitchigin, Larisa Safina, and Manuel Mazzara.
\newblock Verified type-checker for jolie.
\newblock \url{https://arxiv.org/pdf/1703.05186.pdf}.

\bibitem{Bandura16}
Alexey Bandura, Nikita Kurilenko, Manuel Mazzara, Victor Rivera, Larisa Safina,
  and Alexander Tchitchigin.
\newblock Jolie community on the rise.
\newblock In {\em 9th IEEE International Conference on Service-Oriented
  Computing and Applications, SOCA}, 2016.

\bibitem{smtspec}
Clark Barrett, Aaron Stump, and Cesare Tinelli.
\newblock {The SMT-LIB Standard. Version 2.0 }, 2010.

\bibitem{Hoare}
Hoare C.A.R.
\newblock An axiomatic basis for computer programming.
\newblock {\em Communications of the ACM}, 12:576--583, 1969.

\bibitem{Cardelli:1996}
Luca Cardelli.
\newblock Type systems.
\newblock {\em ACM Comput. Surv.}, 28(1):263--264, 1996.

\bibitem{Guidi2017}
Manuel Mazzara Fabrizio~Montesi Claudio~Guidi, Ivan~Lanese.
\newblock Microservices: a language-based approach.
\newblock In {\em Present and Ulterior Software Engineering}. Springer, 2017.

\bibitem{MouraB08}
Leonardo de~Moura and Nikolaj Bjørner.
\newblock Z3: An efficient smt solver.
\newblock In {\em Tools and Algorithms for the Construction and Analysis of
  Systems, 14th International Conference, TACAS 2008, Held as Part of the Joint
  European Conferences on Theory and Practice of Software, ETAPS 2008,
  Budapest, Hungary, March 29-April 6, 2008. Proceedings}, volume 4963 of {\em
  Lecture Notes in Computer Science}, pages 337--340. Springer, 2008.

\bibitem{Dragoni2017b}
N.~Dragoni, I.~Lanese, S.~T. Larsen, M.~Mazzara, R.~Mustafin, and L.~Safina.
\newblock Microservices: How to make your application scale.
\newblock In {\em A.P. Ershov Informatics Conference (the PSI Conference
  Series, 11th edition)}. Springer, 2017.

\bibitem{DDLM2017}
Nicola Dragoni, Schahram Dustdar, Stephan~T. Larse, and Manuel Mazzara.
\newblock Microservices: Migration of a mission critical system.
\newblock \url{https://arxiv.org/abs/1704.04173}.

\bibitem{Dragoni2017}
Nicola Dragoni, Saverio Giallorenzo, Alberto Lluch-Lafuente, Manuel Mazzara,
  Fabrizio Montesi, Ruslan Mustafin, and Larisa Safina.
\newblock Microservices: yesterday, today, and tomorrow.
\newblock In Bertrand Meyer and Manuel Mazzara, editors, {\em Present and
  Ulterior Software Engineering}. Springer, 2017.

\bibitem{Gmehlich2013}
Rainer Gmehlich, Katrin Grau, Felix Loesch, Alexei Iliasov, Michael Jackson,
  and Manuel Mazzara.
\newblock Towards a formalism-based toolkit for automotive applications.
\newblock {\em 2013 1st FME Workshop on Formal Methods in Software Engineering
  (FormaliSE)}, pages 36--42.

\bibitem{LiquidHaskell}
Ranjit Jhala.
\newblock {Liquid Haskell}.
\newblock \url{http://goto.ucsd.edu/~rjhala/liquid/haskell/blog/about/}.

\bibitem{mackenzie2006}
M.C. MacKenzie et~al.
\newblock Reference model for service oriented architecture 1.0.
\newblock {\em OASIS Standard}, 12, 2006.

\bibitem{Nalin2016}
Ilaria~Baroni Marco~Nalin and Manuel Mazzara.
\newblock A holistic infrastructure to support elderlies' independent living.
\newblock {\em Encyclopedia of E-Health and Telemedicine, IGI Global}, 2016.

\bibitem{MazzaraADB11}
M.~Mazzara, F.~Abouzaid, N.~Dragoni, and A.~Bhattacharyya.
\newblock Design, modelling and analysis of a workflow reconfiguration.
\newblock In {\em Proceedings of the International Workshop on Petri Nets and
  Software Engineering, Newcastle upon Tyne, UK, June 20-21, 2011}, pages
  10--24, 2011.

\bibitem{Mazzara:phd}
Manuel Mazzara.
\newblock {\em Towards Abstractions for Web Services Composition}.
\newblock Ph.{D}. thesis, University of Bologna, 2006.

\bibitem{MBGDMQN2013}
Manuel Mazzara, Luca Biselli, Pier~Paolo Greco, Nicola Dragoni, Antonio
  Marraffa, Nafees Qamar, and Simona de~Nicola.
\newblock {\em Social networks and collective intelligence: a return to the
  agora}.
\newblock IGI Global, 2013.

\bibitem{mustafin2017}
Manuel Mazzara, Ruslan Mustafin, Larisa Safina, and Ivan Lanese.
\newblock Towards microservices and beyond: An incoming paradigm shift in
  distributed computing.
\newblock \url{https://arxiv.org/pdf/1610.01778.pdf}.

\bibitem{milnerCCS}
Robin Milner.
\newblock {\em Communication and concurrency}.
\newblock Prentice Hall International (UK) Ltd., 1995.

\bibitem{Milner1999}
Robin Milner.
\newblock {\em Communicating and Mobile Systems: The \&Pgr;-calculus}.
\newblock Cambridge University Press, New York, NY, USA, 1999.

\bibitem{MGZ14}
Fabrizio Montesi, Claudio Guidi, and Gianluigi Zavattaro.
\newblock {Service-Oriented Programming with Jolie}.
\newblock In {\em Web Services Foundations}, pages 81--107. Springer, 2014.

\bibitem{N15}
S.~Newman.
\newblock {\em Building microservices}.
\newblock O'Reilly Media, Inc., 2015.

\bibitem{nielsen}
Julie~Meinicke Nielsen.
\newblock {A Type System for the Jolie Language}.
\newblock Master's thesis, Technical University of Denmark, 2013.

\bibitem{agda}
Chalmers~University of~Technology. Accessed December~2016.
\newblock {Agda}.
\newblock \url{http://wiki.portal.chalmers.se/agda/pmwiki.php}.

\bibitem{Rice53}
Henry~Gordon Rice.
\newblock Classes of recursively enumerable sets and their decision problems.
\newblock {\em Trans. Amer. Math. Soc.}, 74:358--366, 1953.

\bibitem{Rondon2008}
Patrick~M. Rondon, Ming Kawaguci, and Ranjit Jhala.
\newblock Liquid types.
\newblock {\em SIGPLAN Not.}, 43(6):159--169, June 2008.

\bibitem{Safina2016}
Larisa Safina, Manuel Mazzara, Fabrizio Montesi, and Victor Rivera.
\newblock Data-driven workflows for microservices (genericity in jolie).
\newblock In {\em Proc. of The 30th IEEE International Conference on Advanced
  Information Networking and Applications (AINA), 2016}.

\bibitem{Salikhov2016b}
Dilshat Salikhov, Kevin Khanda, Kamill Gusmanov, Manuel Mazzara, and Nikolaos
  Mavridis.
\newblock Jolie good buildings: Internet of things for smart building
  infrastructure supporting concurrent apps utilizing distributed
  microservices.
\newblock In {\em Proceedings of the 1st International conference on Convergent
  Cognitive Information Technologies}, pages 48--53, 2016.

\bibitem{Salikhov2016a}
Dilshat Salikhov, Kevin Khanda, Kamill Gusmanov, Manuel Mazzara, and Nikolaos
  Mavridis.
\newblock Microservice-based iot for smart buildings.
\newblock In {\em Proceedings of the 31st International Conference on Advanced
  Information Networking and Applications Workshops (WAINA)}, 2017.

\bibitem{Tchitchigin16}
Alexander Tchitchigin, Larisa Safina, Manuel Mazzara, Mohamed Elwakil, Fabrizio
  Montesi, and Victor Rivera.
\newblock Refinement types in jolie.
\newblock In {\em Spring/Summer Young Researchers Colloquium on Software
  Engineering, SYRCoSE}, 2016.

\bibitem{Vallverdu16}
Jordi Vallverd\'u, Max Talanov, Salvatore Distefano, Manuel Mazzara, Alexander
  Tchitchigin, and Ildar Nurgaliev.
\newblock A cognitive architecture for the implementation of emotions in
  computing systems.
\newblock {\em Biologically Inspired Cognitive Architectures}, 15(Supplement
  C):34 -- 40, 2016.

\end{thebibliography}

\end{document}